\documentclass[journal=jpcafh,manuscript=article]{achemso}
\usepackage[utf8]{inputenc}
\usepackage[T1]{fontenc}
\usepackage[version=3]{mhchem}
\usepackage{soul}
\usepackage{bm}
\usepackage{threeparttable}
\usepackage{color}


\date{\today}
\author{Paul L. Houston}
\affiliation{Department of Chemistry and Chemical Biology, Cornell University, Ithaca, New York
14853, U.S.A. and Department of Chemistry and Biochemistry, Georgia Institute of
Technology, Atlanta, Georgia 30332, U.S.A}
\email{plh2@cornell.edu}
\author{Apurba Nandi}
\author{Joel M. Bowman}
\email{jmbowma@emory.edu}
\affiliation[Emory University]
{Cherry L. Emerson Center for Scientific Computation and Department of Chemistry, Emory University, Atlanta, Georgia 30322, USA}

\title{A Machine Learning Approach for Rate Constants III: Application to the Cl($^2$P)+\ce{CH4}$\rightarrow$\ce{CH3}+HCl Reaction}

\begin{document}
\newpage

\begin{abstract}
The temperature dependence of the thermal rate constant for the reaction Cl($^3$P) + \ce{CH4} $\rightarrow$ HCl + \ce{CH3} is calculated using a Gaussian Process machine learning (ML) approach to train on and predict thermal rate constants over a large temperature range. Following procedures developed in two previous reports, we use a training dataset of approximately 40 reaction/potential surface combinations, each of which is used to calculate the corresponding data base of rate constant at approximately eight temperatures.  For the current application, we train on the entire dataset and then predict the temperature dependence of the title reaction employing a ``split'' dataset for correction at low and high temperatures to capture both tunneling and recrossing.  The results are an improvement on recent RPMD calculations compared to accurate quantum ones, using the same high-level ab initio potential energy surface. Both tunneling at low temperatures and recrossing at high temperatures are observed to influence the rate constants. Recrossing effects, which are not described by TST and even sophisticated tunneling corrections, do appear in experiment at temperatures above around 600 K. The ML results describe these effects and in fact merge at 600 K with RPMD results (which can describe recrossing), and both are close to experiment at the highest experimental temperatures.
\end{abstract}
\flushbottom
\maketitle

\thispagestyle{empty}
\newpage
\section{Introduction}

Svante Arrhenius nearly turned away from chemistry twice during his early studies because of poor performance, but he went on to earn a Nobel Prize in the field for his work on electrolytic solutions.  Of more relevance to the current topic is his paper in 1889 on the temperature dependence of reaction rates,\cite{Arrhenius1889} where he found that $k(T)=Ae^{-E_{a}/kT}$ where $A$ is a constant, $T$ is the Kelvin temperature, and $E_a$ is an activation energy. A major subsequent advance came from the development in 1935 by Henry Eyring, and by Merideth Evans and Michael Polyani of transition state theory (TST),\cite{Eyring1935,EvansPolanyi1935,LaidlerKing1983} which associated the activation energy with the location of a dividing surface that separated the reactants from the products in the space of energy vs. configuration. For reactions that have a saddle point between reactants and products, the dividing surface typically includes the saddle point.  The constant $A$ is associated in TST with a ratio of partition functions.  

The near-linear relationship between the log of the rate constant with $(1/T)$ predicted by both the Arrhenius equation and TST is actually not observed for many reactions, particularly at low temperatures.  This is because quantum mechanical tunneling is found to be important at low temperatures and results in an increase in the rate constant from the linear prediction. Early proposals for including tunneling effects in the theory of chemical reactions were based on one-dimensional potentials. Of particular relevance to what follows in this paper is Eckart's early work,\cite{eck1930} which continues to be used,\cite{jh1961,jh1962,Barker:2012aa} because it gives an analytical and often accurate result. 

However, to get agreement with exact quantum methods or with experiment, we need more sophisticated quantum mechanical (QM) approaches, even though they are more computationally demanding.  These methods have mostly been applied to collinear reactions, as summarized in a useful compilation by Allison and Truhlar.\cite{at1998}  Three-dimensional calculations, notably for the H + \ce{H2} reaction,\cite{skh31976} have also been studied, and these demonstrate shortcomings of the one-dimensional Eckart model that include, for example, recrossing of the TST dividing surface, corner cutting tunneling paths, and vibrationally adiabatic effects.  To incorporate these effects, a variety of QM approaches have been employed, including instanton methods,\cite{miller75,insthch4rg} VPT2-based semiclassical transition state theory,\cite{MILLER199062,HERNANDEZ1993129,Nguyen:2011aa,Barker:2012aa,Wagner:2013aa,clary18} reduced-dimensionality quantum methods,\cite{bowman1985,bowmanwag,jmb1991,Althorpe2003,claryhalk} a large-curvature, corner-cutting reaction path,\cite{mc1977,at1998}  ring polymer molecular dynamics (RPMD),\cite{rpmd05,rpmdh3} and ring polymer instanton methods.\cite{rich18}  Exact direct methods have also been developed to obtain the rate constant\cite{millerexact83}, for example using multi-configuration time-dependent Hartree (MCTDH) theory, at least for zero total angular momentum,\cite{manthhch4,manth12} and then augmented by $J$-shifting\cite{jmb1991}.    

Until recently, none of these methods had been applied to the important Cl($^3P_{3/2}$) + \ce{CH4} $\rightarrow$ HCl + \ce{CH3} reaction, a reaction important in assessing the effect of chlorofluorocarbons in stratospheric chemistry. Many experimental measurements are available, mostly in very good agreement with one another.  Collectively these measurements span a large temperature range from roughly 180 to more than 1000 K. We will refer to these by the names of the first authors of the reports: Clyne,\cite{Clyne1973} Manning,\cite{Manning1977} Michael,\cite{Michael1977} Whytock,\cite{Whytock1977} Keyser,\cite{Keyser1978} Zahniser,\cite{Zahniser1978} Ravishankara,\cite{Ravishankara1980} Heneghan,\cite{Heneghan1981} Seeley,\cite{Seeley1996} Pilgrim,\cite{Pilgrim1997} and Bryukov.\cite{Bryukov2002}.  High-level, CCSD(T)-based potential energy surfaces for this reaction have been reported by Czakó and Bowman (CB),\cite{CzakoBowman2011} and, quite recently, by Li and Liu (LL).\cite{LL2020}  Substantial theoretical investigations have also  been reported. Barker, Nguyen, and Stanton\cite{BarkerNguyenStanton2012} have reported kinetic isotope effects in the reaction using \textit{ab initio} semiclassical transition state theory, while Georgievskii and Klippenstein (GK)\cite{GK2021} have recently studied how tunneling is affected by the non-separability of the reaction coordinate from other modes. The approach of the latter report provides a method that substantially improves on the one-dimensional estimation of tunneling; the comparison with experiment in that paper was limited to temperatures below 300 K, where agreement is very good. We return to this recent work in the Discussion section below.  Li and Liu\cite{LL2020} used their potential to perform RPMD calculations for  estimation of the temperature dependent rate constant, though over a limited range and with only modest agreement with experiment.  Subsequently, a ``first principles'' theory for the reaction was recently reported by Hoppe and Manthe (HM),\cite{HM2022} who used the LL potential in a fully QM method to calculate the rate constant as a function of temperature up to 540 K. Going to higher temperatures was evidently not computationally feasible. Their results are essentially in exact agreement with the experimental data up to that temperature.  Thus, from the perspective of theory, this reaction has acquired an especially important role for both the potential energy surface and the reaction dynamics/rate constant theory.

One aspect of this reaction that was not directly addressed by these theoretical approaches is the extent of recrossing at high temperature.  Recall that recrossing refers to the reduction of the reaction flux predictions of TST, which, by assumption, ignores recrossing. Recrossing is due to multidimensional dynamical effects where flux that crosses the transition state dividing surface recrosses but back in the direction of reactants. We return to this effect here where we apply our Machine Learning (ML) approach to obtain bimolecular thermal rate constants over a large temperature range.\cite{houston19,NandiOHCl2020} The motivation for this approach was given in detail in those papers and so is just briefly reviewed below. Other ML methods have subsequently been reported for reaction rates by Valleau and co-workers\cite{valleau2020,Valleau2021}, activation energies in an article by Lewis-Atwell, Townsend and Grayson,\cite{Grayson2021} and for activation energies and reaction mechanisms in publications by Green and colleagues.\cite{Green2019,Green2020}  

In our first paper\cite{houston19} we proposed using ML to find a correction to the Eckart tunneling correction to TST.  The great appeal of the Eckart correction is two-fold. First, it is often quite accurate and second it requires no more information than is needed for a standard TST calculation.  Of course this is not an exact correction and more accurate methods (though short of ``exact'') are available, as mentioned above.  However, these methods require substantially more information about the potential energy surface (PES), ranging from an expansion of the PES around the saddle-point beyond the harmonic expansion needed in TST to a global PES needed in RPMD and the MCTDH quantum calculations.  Our goal was to explore whether ML can be applied to correct the Eckart correction and achieve a substantial improvement in accuracy at virtually no additional cost and with no additional information needed about the PES.  The ML approach we took followed the standard protocol of training the desired quantity on a known database. For us the quantity is the correction to the Eckart tunneling correction and the database is a large set of (ideally) exact quantum rate constant calculations for a variety of chemical reactions. We discuss below the database we used. For the ML model we used Gaussian process (GP) regression, which we briefly review in the next section. We used the database of rate constants for 13 reaction/potential surface combinations at typically 8 different temperatures. We created this database by data mining the large compilation of tables of rate constants reported by Allison and Truhlar in 1998. The tables contained TST and exact quantum results for numerous collinear reactions, of which we selected a total of 52.  Also, all parameters needed for an Eckart tunneling correction were given.  From the TST, Eckart correction and exact results the correction to Eckart were calculated. Then GP was trained on the 52 reactions and testing was then done on a set of 39 reaction/potential surface combinations. The GP method, when averaged over all test reactions, was within 80\% of the accurate answer, considerably better than TST (330\%) and Eckart corrected TST (ECK) (110\%). The test reactions included 5 one-dimensional symmetric A + BA reaction/surface combinations; 16 one-dimensional asymmetric A + BC reaction/surface combinations; 15 three-dimensional reaction/surface combinations, both symmetric and asymmetric; and three polyatomic reactions: O($^3P$) + \ce{CH4}, H + \ce{CH4}, and HH + OH.  A challenge for future predictions is that the dataset is relatively small. A strategy, then, is to examine other reactions that are studied by accurate methods, to determine to what extent GP is effective in predicting the new rate constant, and then to add the new reaction to the dataset.  In our second paper\cite{NandiOHCl2020} we focused on the O($^3P$) + HCl reaction on the $^3$A$^{\prime}$ and $^3$A$^{\prime\prime}$ potential energy surfaces. In that paper we took a different approach.  We trained on the entire original dataset and the predicted the rate constant of this reaction. As described in detail in that paper we split the entire data set into two datasets, as briefly reviewed in the next section in given in more detail in Supplementary material (SM). Here we simply note that at low temperatures the corrections to the Eckart tunneling correction are larger than one.  This is expected, as the Eckart correction generally underestimates tunneling, especially in the deep tunneling region.  At high temperatures where there is virtually no Eckart correction, i.e., the correction is about unity,  the correction to Eckart is less than one.  This accounts for recrossing and where in general TST overestimates the rate constant. 

In the next section we briefly summarize the computational approach we use.  We then present new predictions of the rate constant for the Cl($^2$P) + \ce{CH4} reaction.  This reaction is a challenge for our ML approach both because its parameters are at the edge of those of the training set and because it brings up possible issues, such as spin-orbit coupling and Van der Waals wells in both the entrance and exit channels. Detailed comparisons are made with the GP results of this work and the results from GK\cite{GK2021} as well as with the QM results from HM\cite{HM2022} and the RPMD results from LL.\cite{LL2020} We then incorporate ``learning'' this reaction into the correction datasets and verify that when used with ML we do get good agreement with the experiment, as expected.

\section{Computational Approach}
 The  specific approach we proposed\cite{houston19} is summarized next. Let $k(T)$ be an exact rate constant calculated over a temperature range and using a given PES, and let $k^{TST}(T)$ be the conventional TST rate constant using the same PES.  The approach we took earlier is to represent the exact rate constant by the equation
\begin{equation}
k(T)=[\kappa_{ECK}(T)k^{TST}(T)]\chi(T),   
\end{equation}

\noindent where $\kappa_{ECK}(T)$ is the Eckart transmission coefficient.\cite{jh1961,jh1962,Brown1981} and $\chi(T)$ is the correction to the Eckart correction to produce agreement with the accurate rate constant. We reasoned that this approach is better than using ML on the exact $k^{TST}(T)$ directly because $\kappa_{ECK}(T)$ is both easy to obtain and generally gives realistic tunneling corrections.  Thus, the hope is that the GP correction to the Eckart transmission coefficient is small and easy to ``learn''.    
To proceed with machine learning of $\chi(T)$ we need the usual three elements, namely a \textit{database} of exact rate constants, a set of \textit{descriptors}, and a general \textit{machine learning method} to train on the database. We described these in detail previously\cite{houston19}, so we just briefly summarize these key elements here.  

Critical for the success of the ML approach are the parameters that characterize a reaction.  A starting choice includes the three parameters that are needed to obtain the Eckart transmission coefficient,\cite{jh1961}  which, in reduced units, are
\begin{align}
&\alpha_1=V_1/\omega_{im},  \\
&\alpha_2=V_2/\omega_{im},  \\
&u^*(T)=\omega_{im}/(0.69307\hspace{.1cm} T).
\end{align}
$V_1$ ($V_2$) is the saddle point barrier height in cm$^{-1}$ relative to the reactants (products), omitting the zero-point energy in both cases, and where the energy of the reactants is zero. In the case of the Cl($^2$P) + \ce{CH4} reaction, the zero of energy is lowered by the spin-orbit stabilization of the Cl($^2$P$_{3/2}$) atom, about 0.84 kcal/mol.  Using transition state (TS) value of 6.97 kcal/mol from the CCSD(T)/CBS value in Table 3 of GK, this gives a value of $V_1$ = (6.97 - (-0.84)) kcal/mol = 2733 cm$^{-1}$.  We, as well as GK and HM, assume no spin-orbit splitting other than the reactants. Thus, again from the CCSD(T)/CBS value in Table 3 of GK,  $V_2$ = (6.97-5.04) kcal/mol = 675.3 cm$^{-1}$. Note that in both the entrance and exit channels we have assumed that the presence of Van der Waals wells does not affect the $V_1$ and $V_2$ parameters used for the Eckart correction. In short, we ignore these wells. We defer comments about the possible effect of these wells to the Discussion section. The value $\omega_{im}$ is the magnitude of saddle point imaginary frequency (in cm$^{-1}$), which, from Table 2 of GK is 1194 cm$^{-1}$.  The value 0.69307 is an energy conversion factor, equal to $k_B/(h c)$, and $T$ is the temperature in Kelvin.  Improvements were obtained by including an additional parameter, the skew angle, $\beta$, which identifies important mass and corner cutting and recrossing effects. In the case of the Cl($^2$P) + \ce{CH4} reaction, this value is $\beta = 17.3 \textdegree = 0.302$ radians.  More details, particularly the values of these parameters for the training dataset, can be found in our recent paper.\cite{houston19}. 
         
As noted previously,\cite{NandiOHCl2020} it is a challenge to obtain high precision in fitting the entire $\chi$ dataset of approximately 400 $\chi$ values because they have a very large range, from 0.2 to 25. As in the O($^3$P) + HCl paper,\cite{NandiOHCl2020} we thus decided to split the data into two clusters in a manner that would permit more precise GP training on each.  The dividing line between the two clusters is at $\chi$ equal to 3.0.  The cluster with $\chi$ greater than 3.0 is denoted the Large-$\chi$ set and the other cluster is denoted Small-$\chi$ set. Figure S1. in the Supplementary Material (SM) shows these $\chi$ groups as a function of $u^*$. 

Having now a set of descriptors, $\alpha_1$, $\alpha_2$, $\omega_{im}$, and $\beta$, as well as a dataset from ref. \citenum{houston19}, we now turn to the learning method.  Gaussian Process (GP) regression is a machine learning method whose goal is to produce a smooth interpolation between known data.\cite{Rasmussen2004} The working equation is given by
\begin{equation}
\chi(\bm{x}) = \textbf{K}_x^\top \textbf{K}^{-1} \boldsymbol\chi_0,
\label{eq:predict}
\end{equation}
where $\bm{x}$ represents the set of descriptors. The known values of $\chi$ are collected in the column vector denoted $\boldsymbol\chi_0$.  $K(\bm{x}_i,\bm{x}_j)$ is the kernel matrix with elements at the database values of $\bm{x}$ and where $\textbf{K}_x = [k(\bm{x}, \bm{x}_1) \  \cdots \  k(\bm{x}, \textbf{x}_N)]^\top $.  In this expression $\bm{x}$ is the value of the descriptors where $\chi$ is to be evaluated. 
A popular choice for the kernel matrix is\cite{Rasmussen2004,krems17,GP-2015-1}
\begin{equation} \label{eq:noise}
K(\bm{x}_i,\bm{x}_j) = \sigma^2 \exp\left( -\frac{d_{ij}^2}{2}\right)+\delta_{ij}\sigma_{noise},
\end{equation}
where $d_{ij}$ is the distance between the two vectors $\bm{x}_i/l$ and $\bm{x}_j/l$, where the hyper-parameters are $l$ and $\sigma$. The length-scale parameter, $l$ can be single length or one that depends on the descriptor; we use the latter.  $\delta_{ij}\sigma_{noise}$ is the noise term that is added to the diagonal of the covariance matrix. In principle, this term is not necessary for fitting, because the data are not noisy.  However, adding noise can avoid ill-conditioning of the matrix, and, more generally enters parametrically into the optimization of the hyper-parameters according to maximization of log-marginal-likelihood .\cite{Cui2016, Rasmussen2004}
\begin{equation}
\log \mathcal{L} = -\frac{1}{2} \left( \textbf{g}^\top\textbf{K}^{-1}\textbf{g} + \log|\textbf{K}| + N\log2\pi \right)
\label{eq:likelyhood}
\end{equation}
 Once the optimal hyper-parameters are determined, the GP model can predict the value of $\chi$ using Eq. \ref{eq:predict}.  In a practical sense, the hyper-parameters govern the smoothness of the interpolation, also known as prediction. 

We performed this GP regression on the Small and Large-$\chi$ datasets using the routines contained in the Python Scikitlearn library, which includes optimization of the hyperparameters.\cite{scikit-learn}  The inputs are the descriptors and the output is the trained $\chi$.  As a reminder, the database for exact $\chi$ values is developed largely from a 1998 compilation of rate constants\cite{at1998} as well as from results from several more recent quantum calculations of rates constants for polyatomic reactions, such as OH + \ce{H2}, H + \ce{CH4}, and O + \ce{CH4}, as described in detail previously.\cite{houston19}

 The Small-$\chi$ data-set has a total 360 data points, and for the work we described below we took all the points to fit the data-set using GP regression. The RMS error of this fitting is 0.14461. The Large-$\chi$ data-set has total 37 data points, which for ML is very small. In order to avoid over-fitting of this limited set, we used a noise value (see Eq. \ref{eq:noise})  of 50, larger than that used for the previously studied O($^3P$) + HCl reaction. Noise values of 20 and 30 gave nearly identical results, as shown in Figure S4. of the SM.

\section{The \ce{Cl($^2$P_{3/2}) + CH4} $\rightarrow$ HCl + \ce{CH3} Reaction}
The  \ce{Cl($^2$P_{3/2}) + CH4} reaction is a challenge for our ML approach, given the range and sparsity of our relatively small dataset. Recall that the dataset is largely based on collinear A+BC reactions, with only a few three-dimensional reactions.  In particular, for this reaction, using the most accurate electronic energies, $\alpha_1 = 2.61$ and $\alpha_2 = 0.645$ so that $\alpha_1 - \alpha_2 = 1.97$. Our complete dataset does have values of $\alpha_1$ greater than and less than this value of $\alpha_1$; however, the smallest value of $\alpha_2$ in our dataset is 0.70.  And in that case there are only six total entries with this value of $\alpha_2$.  Fortunately, for these entries $\alpha_1$ = 2.94 and the skew angle is 20 deg, which is close to skew angle, $\beta$, of 17.3 deg. Our data set for skew angle ranges from 17 to 89 deg.   The value of $\beta$ for the 
\ce{Cl($^2$P_{3/2}) + CH4} $\rightarrow$ HCl + \ce{CH3} reaction is 17.3\textdegree, near the lower limit of the dataset. So the values of the descriptors for this reaction are at the edges of the training dataset.  Thus, it is interesting to see if a strategy such as that used successfully in the O($^3$P) + HCl reaction will work for \ce{Cl($^2$P_{3/2}) + CH4}.

Then too, there is the question of when to use the results of the GP calculation on the Large-$\chi$ set and when to use those of the Small-$\chi$ dataset. Only at the lowest temperatures are the $u^*$ values only found in the former, so one must make a transition between using the two sets that does not result in a discontinuity in the slope of the log$_{10}(k(T))$ vs $1000/T$ plot. Here, we use a transition between the two sets based on the probability that a given $u^*$ will be associated with one set rather than the other, as described in Section S-I of the SM along with some details of the GP calculations using each dataset. We used a different strategy in paper 2 where a weighted average of predicted $\chi$ values was used for several temperatures. \cite{NandiOHCl2020}

Table \ref{tab:GPcalculation} provides a summary of the GP-predicted results along with TST and Eckart corrected TST and experiment.  The TST result is calculated either by direct count of the rotational levels and degeneracies or by using classical formulae, with practically identical results for the temperatures listed.  Our TST calculations are in agreement with those of HM and GK, following discussions with HM, as described in Section S-IV of the SM.  The ``accurate'' value of the rate constant is taken as the result of a third-order polynomial fit to the experimental log$_{10}(k(t))$ vs $1000/T$ plot, shown in Fig. S1 in Section S-II.  As mentioned, the Eckart correction $\kappa$ predicts only a factor of ca. 2 correction from the TST rate constant at the lowest temperatures and is otherwise close to unity (i.e., no correction).  The GP calculation gives a further factor of ca. 5.9 correction at the lowest temperature, but falls somewhat short of the GP target factor, 10.2, that would be needed to provide agreement with experiment, i.e., ``accurate''.

\begin{table}[ht]
\caption{Results for GP Calculation}
\label{tab:GPcalculation}
\begin{threeparttable}
\centering
	\begin{tabular*}{\columnwidth}{cccccccc}
	\hline
	\hline\noalign{\smallskip}

  T & u$^*$ &
  TST & Eckart & GP  & $k_{GP}$ & GP & 
    $k_{expt.}$  \\
    (K) & & (this work) & Corrctn. & Corrctn. &  & Target & 
    \\
    & & & $\kappa$ & $\chi$ & & & \\
    \hline
	\noalign{\smallskip}
 150 & 10.00 & 6.89$\times$ 10$^{-17}$ & 1.98 & 5.93 & 8.08$\times$ 10$^{-16}$ & 10.20 & 1.39$\times$ 10$^{-15}$  \\
 200 & 7.53 & 1.44$\times$ 10$^{-15}$ & 1.62 & 2.60 & 6.07$\times$ 10$^{-15}$ & 5.43 & 1.27$\times$ 10$^{-14}$  \\
 300 & 5.02 & 3.48$\times$ 10$^{-14}$ & 1.34 & 0.89 & 4.13$\times$ 10$^{-14}$ & 2.33 & 1.09$\times$ 10$^{-13}$  \\
 400 & 3.77 & 2.01$\times$ 10$^{-13}$ & 1.23 & 0.64 & 1.57$\times$ 10$^{-13}$ & 1.60 & 3.94$\times$ 10$^{-13}$  \\
 500 & 3.01 & 6.49$\times$ 10$^{-13}$ & 1.17 & 0.58 & 4.38$\times$ 10$^{-13}$ & 1.28 & 9.67$\times$ 10$^{-13}$ \\
 600 & 2.51 & 1.55$\times$ 10$^{-12}$ & 1.13 & 0.55 & 9.70$\times$ 10$^{-13}$ & 1.07 & 1.88$\times$ 10$^{-12}$  \\
 700 & 2.15 & 3.06$\times$ 10$^{-12}$ & 1.11 & 0.54 & 1.83$\times$ 10$^{-12}$ & 0.92 & 3.12$\times$ 10$^{-12}$  \\
 800 & 1.88 & 5.34$\times$ 10$^{-12}$ & 1.09 & 0.53 & 3.09$\times$ 10$^{-12}$ & 0.80 & 4.67$\times$ 10$^{-12}$  \\
 900 & 1.67 & 8.53$\times$ 10$^{-12}$ & 1.08 & 0.52 & 4.80$\times$ 10$^{-12}$ & 0.70 & 6.48$\times$ 10$^{-12}$  \\
 950 & 1.59 & 1.05$\times$ 10$^{-11}$ & 1.07 & 0.52 & 5.83$\times$ 10$^{-12}$ & 0.66 & 7.46$\times$ 10$^{-12}$  \\
 1000 & 1.51 & 1.28$\times$ 10$^{-11}$ & 1.07 & 0.51 & 7.00$\times$ 10$^{-12}$ & 0.62 & 8.48$\times$ 10$^{-12}$  \\
 1100 & 1.37 & 1.81$\times$ 10$^{-11}$ & 1.06 & 0.51 & 9.73$\times$ 10$^{-12}$ & 0.55 & 1.06$\times$ 10$^{-11}$  \\
 1200 & 1.26 & 2.46$\times$ 10$^{-11}$ & 1.05 & 0.50 & 1.30$\times$ 10$^{-11}$ & 0.50 & 1.29$\times$ 10$^{-11}$ \\
 1300 & 1.16 & 3.24$\times$ 10$^{-11}$ & 1.05 & 0.50 & 1.69$\times$ 10$^{-11}$ & 0.45 & 1.52$\times$ 10$^{-11}$  \\
 1400 & 1.08 & 4.15$\times$ 10$^{-11}$ & 1.04 & 0.49 & 2.13$\times$ 10$^{-11}$ & 0.41 & 1.76$\times$ 10$^{-11}$  \\
 1500 & 1.00 & 5.19$\times$ 10$^{-11}$ & 1.04 & 0.49 & 2.63$\times$ 10$^{-11}$ & 0.37 & 2.00$\times$ 10$^{-11}$  \\
 1600 & 0.94 & 6.36$\times$ 10$^{-11}$ & 1.04 & 0.48 & 3.18$\times$ 10$^{-11}$ & 0.34 & 2.24$\times$ 10$^{-11}$  \\
 1700 & 0.89 & 7.67$\times$ 10$^{-11}$ & 1.04 & 0.48 & 3.80$\times$ 10$^{-11}$ & 0.31 & 2.47$\times$ 10$^{-11}$  \\
 1800 & 0.84 & 9.10$\times$ 10$^{-11}$ & 1.03 & 0.48 & 4.47$\times$ 10$^{-11}$ & 0.29 & 2.71$\times$ 10$^{-11}$  \\
 1900 & 0.79 & 1.07$\times$ 10$^{-10}$ & 1.03 & 0.47 & 5.19$\times$ 10$^{-11}$ & 0.27 & 2.94$\times$ 10$^{-11}$  \\
 	\hline
	\hline\noalign{\smallskip}
   \end{tabular*}
\end{threeparttable}
\end{table}

\begin{figure}[h!]
    \centering
    \includegraphics[width=1.0\columnwidth]{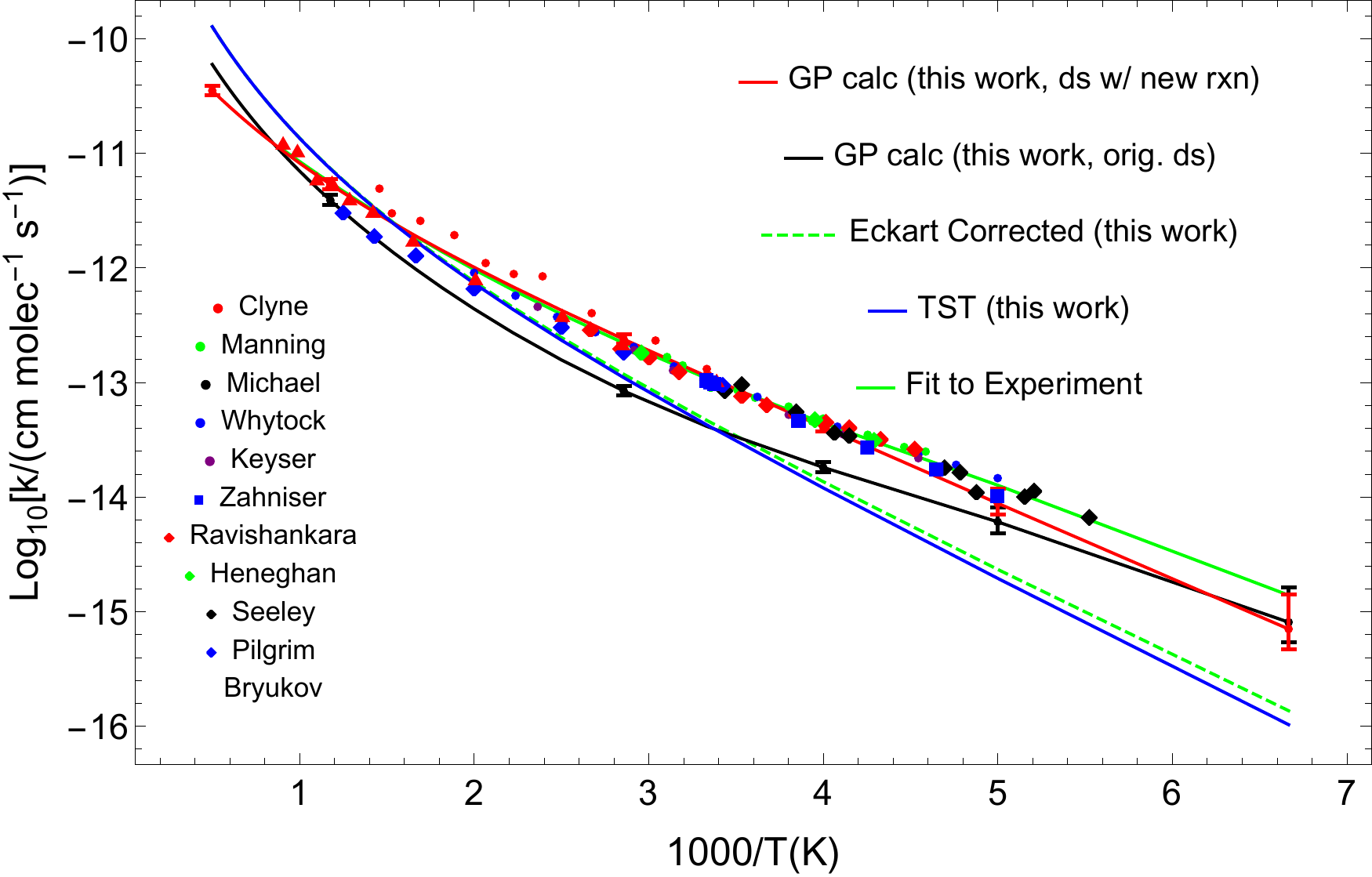}
    \caption{Cl + \ce{CH4} results for the GP method. The method starts with the TST rate constant (solid blue), performs a standard Eckart correction (producing the dashed green curve), and then performs a GP correction based on the original dataset (producing the solid black curve with representative error bars).  If the Cl + \ce{CH4} reaction is included in the dataset, the GP calculation gives the solid red line, with representative error bars.}
    \label{fig:ourresults}
\end{figure}

Figure \ref{fig:ourresults} shows the progression of our calculations.  The blue line gives the TST result, whereas the dashed green line gives the Eckart-corrected TST result. As can be seen, there is very little difference between these two; the Eckart correction is small.  The black line with error bars shows the results of the further correction from the GP calculation, whereas the solid green line gives the result of the fit to the experiment. The GP correction substantially improved the low-temperature rate constants but does not quite provide quantitative agreement with the experiment. (Note that the GP correction uses $\chi$ values from the Large and Small $\chi$ datasets as described in detail in the SM. At the lowest three temperatures of Table \ref{tab:GPcalculation} the contributions from the Large-$\chi$ datasets are 87, 53, and 11 percent, respectively.  At 400 K or above the contributions are less than 3 percent; at 700 K or above they are less thatn 0.5 percent.) Neverthless, there is a major improvement over TST and also the Eckart correction at both low and high temperatures.  At high temperatures  both TST and Eckart corrected TST overestimates the accurate result by about a factors of 2 - 5.  

The representative error bars shown in Fig. \ref{fig:ourresults} are caused by two sources, the uncertainties for the ML predictions using different noise values and the uncertainties for using differing methods for combining the small- and large-$\chi$ data sets.  The former is smaller than the latter.  Fig. S4 shows predictions for different noise values used in the large-$\chi$ set; the curves are close except at the lowest temperatures.  A larger error, shown by the representative error bars in Fig. \ref{fig:ourresults} comes from the latter error source.  The method we used (see Section S-I) was not the only method we tested, and the representative error bars include uncertainty from the different methods.

Of course, one goal of this study is to expand the dataset. The red line with error bars shows the GP result following augmentation of the training dataset with the data for the \ce{Cl($^2$P_{3/2}) + CH4} reaction under investigation.  When the reaction is included in the dataset with the target $\chi$ values, the GP calculate reproduces the experimental data down to $1000/T = 4$, but then still underestimates slightly the needed correction at the lowest temperatures.  We note that the data, the original GP calculation, and the GP calculation made with the augmented training dataset all predict lower rate constants at very high temperatures, when compared to the TST rate constants, correctly capturing the effect of recrossing trajectories.

\section*{Discussion}

Despite the small size of our training dataset, the partial coverage  of it and other challenges posed by the  \ce{Cl($^2$P_{3/2}) + CH4} $\rightarrow$ HCl + \ce{CH3} reaction, the results shown above demonstrate that it is both possible to obtain an good prediction of the temperature dependence of a new reaction and that the prediction is improved substantially when the dataset is expanded by including the new reaction. We now investigate how the GP prediction compares with others. 

\begin{figure}[h!t]
    \centering
    \includegraphics[width=0.85\columnwidth]{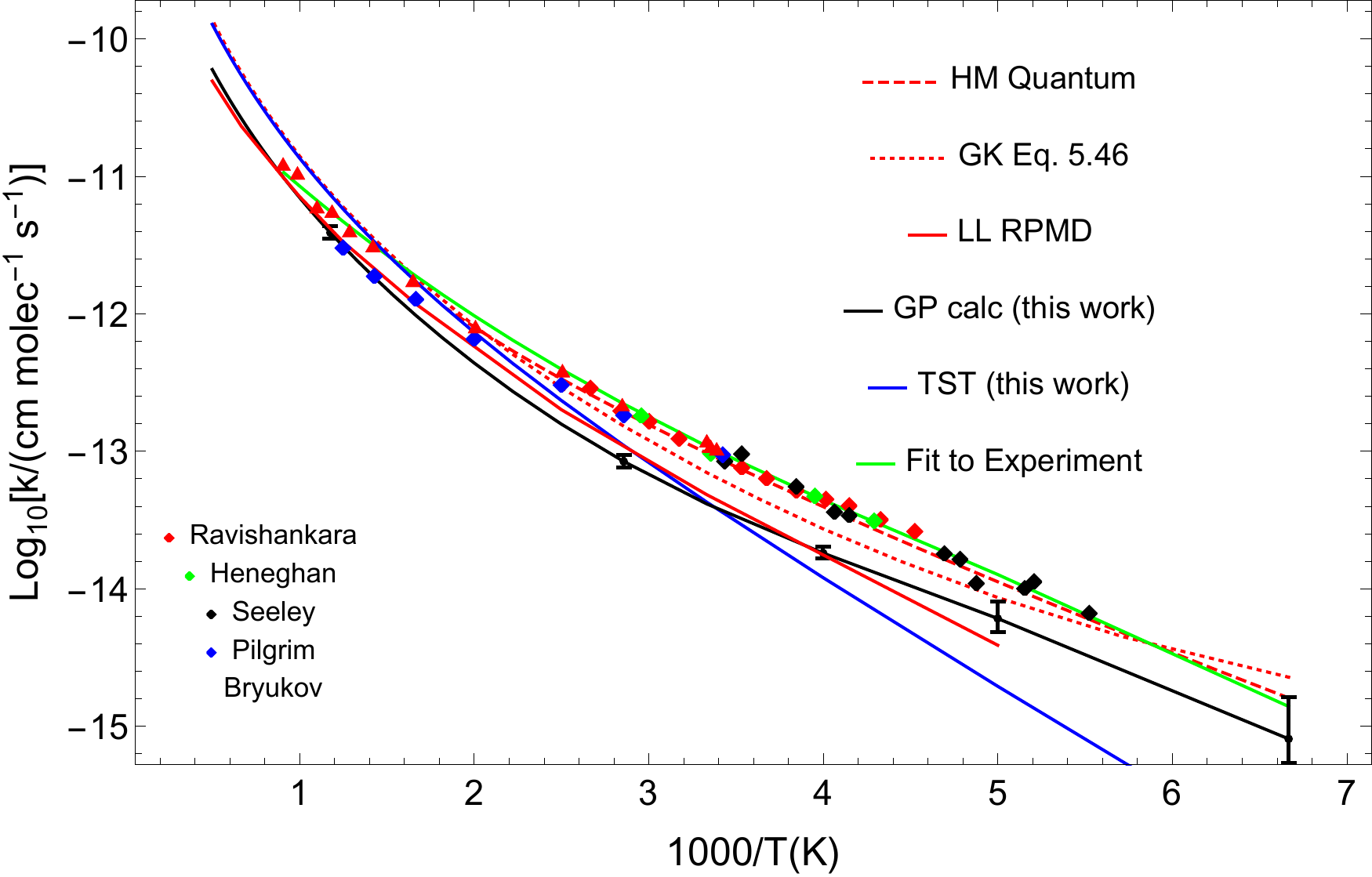}
    \includegraphics[width=0.85\columnwidth]{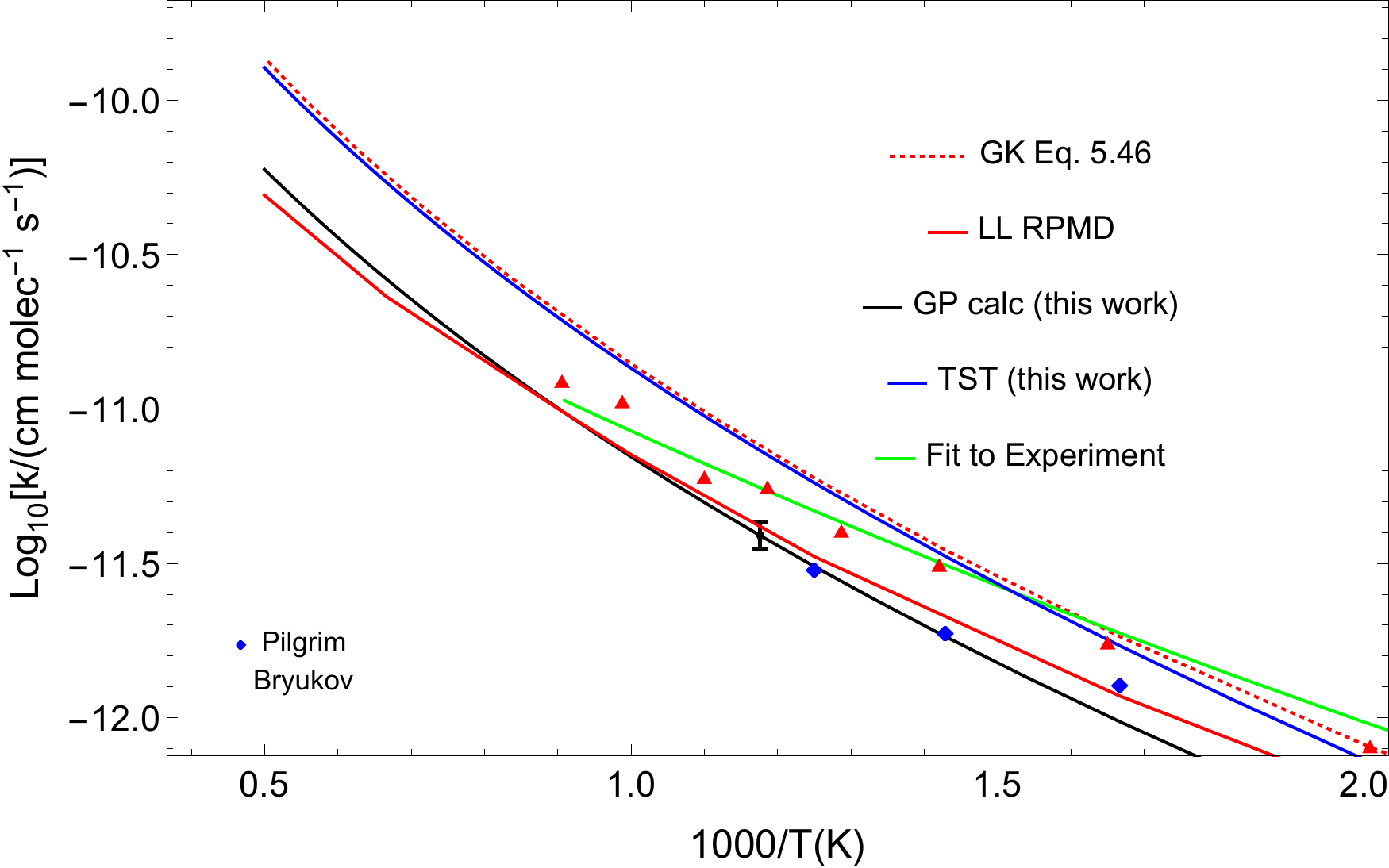}
    \caption{Top Panel: Comparison of results for the Cl + \ce{CH4} reaction showing a sampling of experimental results and the fit to all the experimental data (solid green), the quantum results from ref. \citenum{Heneghan1981} (dashed red), the results from ref. \citenum{GK2021} (dotted red), our GP results  using the original training set (solid black), the RPMD results of ref. \citenum{LL2020} (solid red), and the TST results (solid blue). Bottom Panel: Expansion of the High-T region.}.
    \label{fig:comparison}
\end{figure}

Figure \ref{fig:comparison}. shows a comparison of recent results from several groups.  Liu and Li\cite{LL2020} used a machine-learning approach to develop a potential energy surface for the Cl + \ce{CH4} reaction and then performed RPMD calculations of the rate constant for several temperatures in a limited range.  Their result is shown as the solid red line in the figure, which extends from $1000/T = 0.5-5.0$.  The GP results from this work using the original training set are shown as the solid black line with representative error bars.  The agreement between the two is good, but neither agrees perfectly with the experimental results, a sampling of which is shown along with the green line fit to all the experimental data.  The results from the method of Georgievskii and Klippenstein\cite{GK2021} are shown in the dotted red line.  Hoppe and Manthe\cite{HM2022} used the LL PES to perform quantum calculations of the rate constant as a function of temperature, and their results are shown in the dashed red line extending from $1000/T=1.85-6.67$.  Both the GK and HH results are in relatively good agreement with the experiment, with the HM calculation being almost in exact agreement.  

It is interesting to note that our GP results, using the original data set, and so a prediction, are in  slightly in better overall agreement with experiment than the RPMD ones.  The former requires only 4 parameters, a simple program, and a basic dataset; the calculation takes about a minute on a laptop.  The latter calculation requires a full PES, a much more elaborate code, and a substantially longer calculation time.  It is also curious that the RPMD calculation is not in better agreement with full QM calculation of HM, especially since both used the same PES.  RPMD is essentially exact for the H + \ce{CH4} reaction\cite{rpmdch52015} as shown in Figure. S4 of ref. \citenum{houston19}, as well as for the O + \ce{CH4} reaction\cite{rpmdclch4} as shown in Figure. 3 of ref.\citenum{houston19} and for the O($^3$P) + HCl reaction\cite{rpmd2019} as shown in Figure. 9 of ref. \citenum{NandiOHCl2020}. Whatever the answer is to this puzzle, it is clear that the GP method is effective in making a substantial correction to the TST or Eckart corrected TST rate constants, as we have seen in Figures. \ref{fig:ourresults} and \ref{fig:comparison}.

Another observation is that both our GP result and the data suggest that the rate constant is reduced at high temperatures compared to the TST result, likely because of recrossing.  At first it seems surprising that neither the HM nor GK result captures this effect.  In the HM study, the authors specifically commented that they did not see the trapped oscillatory motion of the H atom due to the heavy-light-heavy mass combination. However, it seems likely from Figure. \ref{fig:comparison} that their quantum calculation was not performed at high enough temperatures to observe this effect; the dashed red line giving their results stops at $1000/T =1.85$ or $T=540$ K,. The approach of GK, while effectively accounting for the tunneling, was not designed to cover recrossing.  The RPMD calculation of LL and our GP calculation do appear to capture the effect, as shown most clearly in the bottom panel of Figure. \ref{fig:comparison}. We do note that the GP predictions using the original datasets overestimates the extent of recrossing at temperatures between around 340 - 600 K.

Finally, we comment briefly on the possible effects of van der Waals wells in this reaction and in general. The importance of these for rate constants was perhaps noted first for the O($^3$P)+HCl reaction.\cite{tiao2002,xieohcl,rpmd2019} The reaction was the focus of a ``stress test'' for RPMD \cite{rpmd2019}, and as noted already it was one focus of our paper II in this series.\cite{NandiOHCl2020} The fundamental effect is a quantum one, where resonances in these well enhance the reactivity at energies below the barrier and thus enhance tunneling. This quantum effect is not described in RPMD, and so the absence of this enhanced tunneling was suggested as the reason RPMD underestimates the exact quantum rate constant (on the same PES).\cite{rpmd2019}  We don't know if these effects are involved in the present reaction, which as noted does have van der Waals wells, and it is beyond the scope of this paper to investigate this issue. However, it is interesting to speculate on how these wells might be approximately incorporated into a 1d tunneling correction such as the Eckart correction. One of us presented a simple heuristic 1d model that treats the resonances as metastable states that increase tunneling through a 1d barrier.\cite{bowman05} That might be a first step towards such theory.

\section{Conclusions}

The Gaussian Process regression method for machine learning of thermal rate constants developed previously\cite{houston19,NandiOHCl2020} has been shown here to provide a good estimate of the temperature dependence of the rate constant for the  \ce{Cl($^2$P_{3/2}) + CH4} $\rightarrow$ HCl + \ce{CH3} reaction. The results suggest both that tunneling is important at temperatures below about 250 K and that recrossing is important at temperatures above about 500 K. The results are in fairly good agreement with RPMD calculations by Liu and Li\cite{LL2020}, but neither method is as accurate in predicting the experimental results as the predictions of the method used by Georgievskii and Klippenstein\cite{GK2021} or, especially, the quantum calculations of Hoppe and Manthe,\cite{HM2022} in the deep tunneling regime. However, the present ML and RMPD results are closer to experiment at higher temperatures, where recrossing becomes signficant.  When the new reaction is included in the training dataset, the GP results are in agreement with experiment to within the error limits, suggesting that the new training set will be more effective in predicting rates for new three-dimensional systems in the future. We note that the original training basis set\cite{at1998} was based predominantly on the results of one-dimensional reaction calculations.  As more complex reactions are added to the dataset, we can expect improvements in the prediction of other reactions such as the Cl + \ce{CH4} one.

\section{Acknowledgments}
JMB  thanks NASA (grant 80NSSC20K0360) for financial support. We thank Yang Liu and Jun Li for providing their RPMD results, Hannes Hoppe and Uwe Manthe for providing their quantum and TST results and for extensive discussions, and Yuri Georgievskii and Stephen Klippenstein for providing  the results of their Eq. 5.46.   
\newpage
\section{TOC Figure}
\begin{figure}[h!]
    \centering
    \includegraphics[width=0.6\columnwidth]{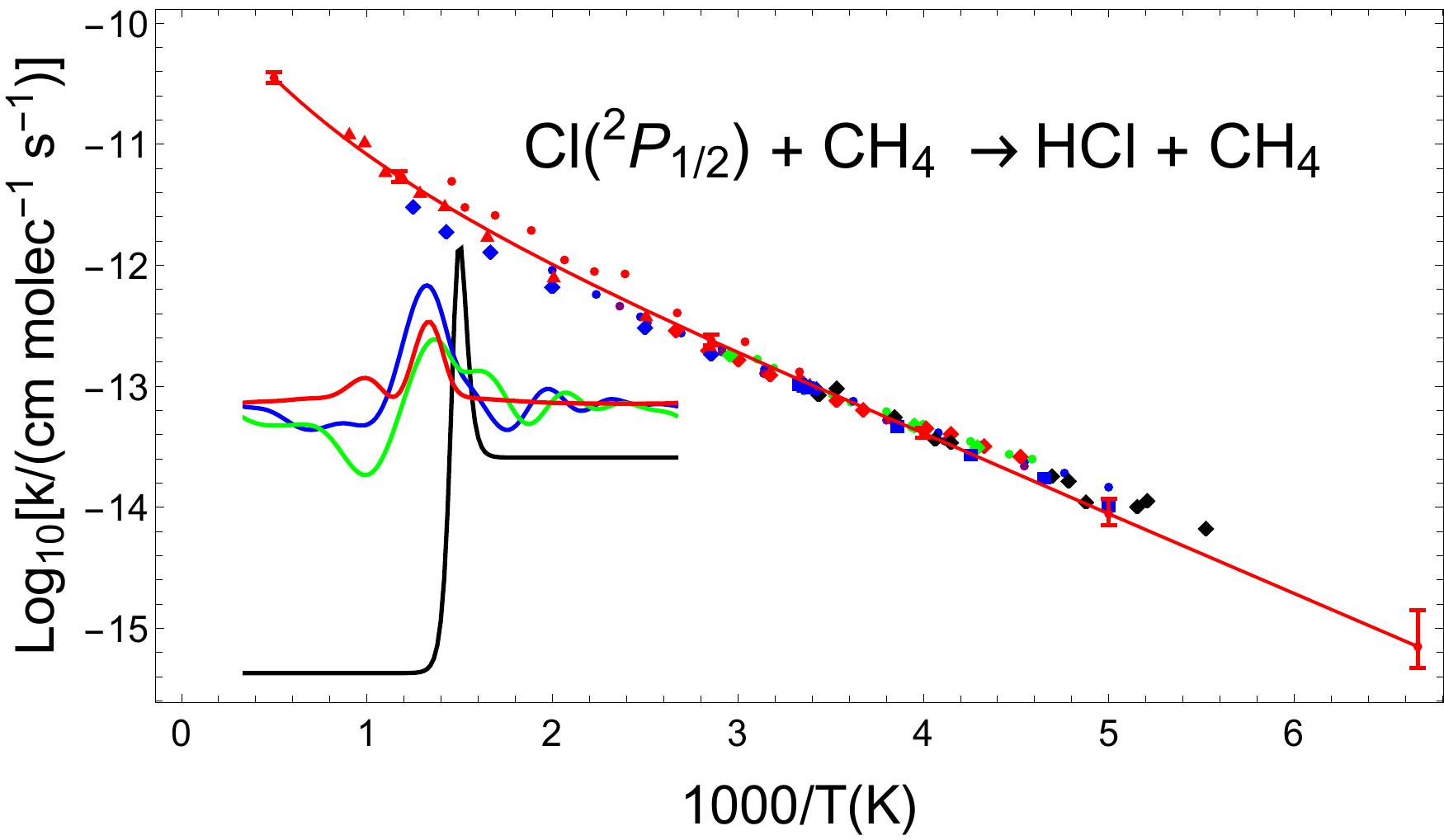}
\end{figure}
\bibliography{refs}

\providecommand{\latin}[1]{#1}
\makeatletter
\providecommand{\doi}
  {\begingroup\let\do\@makeother\dospecials
  \catcode`\{=1 \catcode`\}=2 \doi@aux}
\providecommand{\doi@aux}[1]{\endgroup\texttt{#1}}
\makeatother
\providecommand*\mcitethebibliography{\thebibliography}
\csname @ifundefined\endcsname{endmcitethebibliography}
  {\let\endmcitethebibliography\endthebibliography}{}
\begin{mcitethebibliography}{65}
\providecommand*\natexlab[1]{#1}
\providecommand*\mciteSetBstSublistMode[1]{}
\providecommand*\mciteSetBstMaxWidthForm[2]{}
\providecommand*\mciteBstWouldAddEndPuncttrue
  {\def\EndOfBibitem{\unskip.}}
\providecommand*\mciteBstWouldAddEndPunctfalse
  {\let\EndOfBibitem\relax}
\providecommand*\mciteSetBstMidEndSepPunct[3]{}
\providecommand*\mciteSetBstSublistLabelBeginEnd[3]{}
\providecommand*\EndOfBibitem{}
\mciteSetBstSublistMode{f}
\mciteSetBstMaxWidthForm{subitem}{(\alph{mcitesubitemcount})}
\mciteSetBstSublistLabelBeginEnd
  {\mcitemaxwidthsubitemform\space}
  {\relax}
  {\relax}

\bibitem[Arrhenius(1889)]{Arrhenius1889}
Arrhenius,~S. Über die Reaktionsgeschwindigkeit bei der Inversion von
  Rohrzucker durch Säuren. \emph{Z. Phys. Chem.} \textbf{1889}, \emph{4U},
  226--248\relax
\mciteBstWouldAddEndPuncttrue
\mciteSetBstMidEndSepPunct{\mcitedefaultmidpunct}
{\mcitedefaultendpunct}{\mcitedefaultseppunct}\relax
\EndOfBibitem
\bibitem[Eyring(1935)]{Eyring1935}
Eyring,~H. The Activated Complex in Chemical Reactions. \emph{J. Chem. Phys.}
  \textbf{1935}, \emph{3}, 107--115\relax
\mciteBstWouldAddEndPuncttrue
\mciteSetBstMidEndSepPunct{\mcitedefaultmidpunct}
{\mcitedefaultendpunct}{\mcitedefaultseppunct}\relax
\EndOfBibitem
\bibitem[Evans and Polanyi(1935)Evans, and Polanyi]{EvansPolanyi1935}
Evans,~M.~G.; Polanyi,~M. Some applications of the transition state method to
  the calculation of reaction velocities, especially in solution. \emph{Trans.
  Faraday Soc.} \textbf{1935}, \emph{31}, 875--894\relax
\mciteBstWouldAddEndPuncttrue
\mciteSetBstMidEndSepPunct{\mcitedefaultmidpunct}
{\mcitedefaultendpunct}{\mcitedefaultseppunct}\relax
\EndOfBibitem
\bibitem[Laidler and King(1983)Laidler, and King]{LaidlerKing1983}
Laidler,~K.~J.; King,~M.~C. Development of transition-state theory. \emph{J.
  Phys. Chem.} \textbf{1983}, \emph{87}, 2657--2664\relax
\mciteBstWouldAddEndPuncttrue
\mciteSetBstMidEndSepPunct{\mcitedefaultmidpunct}
{\mcitedefaultendpunct}{\mcitedefaultseppunct}\relax
\EndOfBibitem
\bibitem[Eckart(1930)]{eck1930}
Eckart,~C. The Penetration of a Potential Barrier by Electrons. \emph{Phys.
  Rev.} \textbf{1930}, \emph{35}, 1303--1309\relax
\mciteBstWouldAddEndPuncttrue
\mciteSetBstMidEndSepPunct{\mcitedefaultmidpunct}
{\mcitedefaultendpunct}{\mcitedefaultseppunct}\relax
\EndOfBibitem
\bibitem[Johnston and Rapp(1961)Johnston, and Rapp]{jh1961}
Johnston,~H.~S.; Rapp,~D. Large Tunnelling Corrections in Chemical Reaction
  Rates. II. \emph{J. Am. Chem. Soc.} \textbf{1961}, \emph{83}, 1--9\relax
\mciteBstWouldAddEndPuncttrue
\mciteSetBstMidEndSepPunct{\mcitedefaultmidpunct}
{\mcitedefaultendpunct}{\mcitedefaultseppunct}\relax
\EndOfBibitem
\bibitem[Johnston and Heicklen(1962)Johnston, and Heicklen]{jh1962}
Johnston,~H.~S.; Heicklen,~J. Tunneling Corrections for Unsymmetrical Eckart
  Potential Energy Barriers. \emph{J. Phys. Chem.} \textbf{1962}, \emph{66},
  532--533\relax
\mciteBstWouldAddEndPuncttrue
\mciteSetBstMidEndSepPunct{\mcitedefaultmidpunct}
{\mcitedefaultendpunct}{\mcitedefaultseppunct}\relax
\EndOfBibitem
\bibitem[Barker \latin{et~al.}(2012)Barker, Nguyen, and Stanton]{Barker:2012aa}
Barker,~J.~R.; Nguyen,~T.~L.; Stanton,~J.~F. Kinetic Isotope Effects for
  \uppercase{C}l + \uppercase{CH$_4$} $\rightarrow$ \uppercase{HC}l +
  \uppercase{CH$_3$} Calculated Using Ab Initio Semiclassical Transition State
  Theory. \emph{J. Phys. Chem. A.} \textbf{2012}, \emph{116}, 6408--19\relax
\mciteBstWouldAddEndPuncttrue
\mciteSetBstMidEndSepPunct{\mcitedefaultmidpunct}
{\mcitedefaultendpunct}{\mcitedefaultseppunct}\relax
\EndOfBibitem
\bibitem[Allison and Truhlar(1998)Allison, and Truhlar]{at1998}
Allison,~T.~C.; Truhlar,~D.~G. In \emph{Modern Methods for Multidimensional
  Dynamics Computations in Chemistry}; Thompson,~D.~L., Ed.; World Scientific:
  Singapore, 1998; pp 618--712\relax
\mciteBstWouldAddEndPuncttrue
\mciteSetBstMidEndSepPunct{\mcitedefaultmidpunct}
{\mcitedefaultendpunct}{\mcitedefaultseppunct}\relax
\EndOfBibitem
\bibitem[Schatz and Kuppermann(1976)Schatz, and Kuppermann]{skh31976}
Schatz,~G.~C.; Kuppermann,~A. Quantum Mechanical Reactive Scattering for
  Three-Dimensional Atom Plus Diatom Systems. II. Accurate Cross Sections for
  \uppercase{H+H$_2$}. \emph{J. Chem. Phys.} \textbf{1976}, \emph{65},
  4668--4692\relax
\mciteBstWouldAddEndPuncttrue
\mciteSetBstMidEndSepPunct{\mcitedefaultmidpunct}
{\mcitedefaultendpunct}{\mcitedefaultseppunct}\relax
\EndOfBibitem
\bibitem[Miller(1975)]{miller75}
Miller,~W.~H. Semiclassical Limit of Quantum Mechanical Transition State Theory
  for Nonseparable Systems. \emph{J. Chem. Phys.} \textbf{1975}, \emph{62},
  1899--1906\relax
\mciteBstWouldAddEndPuncttrue
\mciteSetBstMidEndSepPunct{\mcitedefaultmidpunct}
{\mcitedefaultendpunct}{\mcitedefaultseppunct}\relax
\EndOfBibitem
\bibitem[Zhao \latin{et~al.}(2004)Zhao, Yamamoto, and Miller]{insthch4rg}
Zhao,~Y.; Yamamoto,~T.; Miller,~W.~H. Path Integral Calculation of Thermal Rate
  Constants within the Quantum Instanton Approximation: Application to the
  \uppercase{H+CH$_4$} $\rightarrow$ \uppercase{H$_2$+CH$_3$} Hydrogen
  Abstraction Reaction in Full Cartesian Space. \emph{J. Chem. Phys.}
  \textbf{2004}, \emph{120}, 3100--3107\relax
\mciteBstWouldAddEndPuncttrue
\mciteSetBstMidEndSepPunct{\mcitedefaultmidpunct}
{\mcitedefaultendpunct}{\mcitedefaultseppunct}\relax
\EndOfBibitem
\bibitem[Miller \latin{et~al.}(1990)Miller, Hernandez, Handy, Jayatilaka, and
  Willetts]{MILLER199062}
Miller,~W.~H.; Hernandez,~R.; Handy,~N.~C.; Jayatilaka,~D.; Willetts,~A. Ab
  initio Calculation of Anharmonic Constants for a Transition State, with
  Application to Semiclassical Transition State Tunneling Probabilities.
  \emph{Chem. Phys. Lett.} \textbf{1990}, \emph{172}, 62 -- 68\relax
\mciteBstWouldAddEndPuncttrue
\mciteSetBstMidEndSepPunct{\mcitedefaultmidpunct}
{\mcitedefaultendpunct}{\mcitedefaultseppunct}\relax
\EndOfBibitem
\bibitem[Hernandez and Miller(1993)Hernandez, and Miller]{HERNANDEZ1993129}
Hernandez,~R.; Miller,~W.~H. Semiclassical Transition State Theory. A New
  Perspective. \emph{Chem. Phys. Lett.} \textbf{1993}, \emph{214}, 129 --
  136\relax
\mciteBstWouldAddEndPuncttrue
\mciteSetBstMidEndSepPunct{\mcitedefaultmidpunct}
{\mcitedefaultendpunct}{\mcitedefaultseppunct}\relax
\EndOfBibitem
\bibitem[Nguyen \latin{et~al.}(2011)Nguyen, Stanton, and Barker]{Nguyen:2011aa}
Nguyen,~T.~L.; Stanton,~J.~F.; Barker,~J.~R. Ab Initio Reaction Rate Constants
  Computed Using Semiclassical Transition-State Theory: \uppercase{HO + H$_2$}
  $\rightarrow$ \uppercase{H$_2$O + H} and Isotopologues. \emph{J. Phys. Chem.
  A.} \textbf{2011}, \emph{115}, 5118--26\relax
\mciteBstWouldAddEndPuncttrue
\mciteSetBstMidEndSepPunct{\mcitedefaultmidpunct}
{\mcitedefaultendpunct}{\mcitedefaultseppunct}\relax
\EndOfBibitem
\bibitem[Wagner(2013)]{Wagner:2013aa}
Wagner,~A.~F. Improved Multidimensional Semiclassical Tunneling Theory.
  \emph{J. Phys. Chem. A.} \textbf{2013}, \emph{117}, 13089--100\relax
\mciteBstWouldAddEndPuncttrue
\mciteSetBstMidEndSepPunct{\mcitedefaultmidpunct}
{\mcitedefaultendpunct}{\mcitedefaultseppunct}\relax
\EndOfBibitem
\bibitem[Clary(2018)]{clary18}
Clary,~D.~C. Spiers Memorial Lecture Introductory Lecture: Quantum Dynamics of
  Chemical Reactions. \emph{Faraday Discuss.} \textbf{2018}, \emph{212},
  9--32\relax
\mciteBstWouldAddEndPuncttrue
\mciteSetBstMidEndSepPunct{\mcitedefaultmidpunct}
{\mcitedefaultendpunct}{\mcitedefaultseppunct}\relax
\EndOfBibitem
\bibitem[Bowman(1985)]{bowman1985}
Bowman,~J.~M. \emph{Adv. Chem. Phys.}; J. Wiley and Sons, Ltd, 1985; pp
  115--167\relax
\mciteBstWouldAddEndPuncttrue
\mciteSetBstMidEndSepPunct{\mcitedefaultmidpunct}
{\mcitedefaultendpunct}{\mcitedefaultseppunct}\relax
\EndOfBibitem
\bibitem[Bowman and Wagner(1986)Bowman, and Wagner]{bowmanwag}
Bowman,~J.~M.; Wagner,~A.~F. In \emph{{The Theory of Chemical Reaction
  Dynamics}}; Clary,~D., Ed.; NATO ASI Series (Series C: Mathematical and
  Physical Sciences); Springer, Dordrecht, 1986; Vol. 170; Chapter 6, pp
  129--164\relax
\mciteBstWouldAddEndPuncttrue
\mciteSetBstMidEndSepPunct{\mcitedefaultmidpunct}
{\mcitedefaultendpunct}{\mcitedefaultseppunct}\relax
\EndOfBibitem
\bibitem[Bowman(1991)]{jmb1991}
Bowman,~J.~M. Reduced Dimensionality Theory of Quantum Reactive Scattering.
  \emph{J. Phys. Chem.} \textbf{1991}, \emph{95}, 129--164\relax
\mciteBstWouldAddEndPuncttrue
\mciteSetBstMidEndSepPunct{\mcitedefaultmidpunct}
{\mcitedefaultendpunct}{\mcitedefaultseppunct}\relax
\EndOfBibitem
\bibitem[Althorpe and Clary(2003)Althorpe, and Clary]{Althorpe2003}
Althorpe,~S.~C.; Clary,~D.~C. Quantum Scattering Calculations on Chemical
  Reactions. \emph{Annu. Rev. Phys. Chem.} \textbf{2003}, \emph{54},
  493--529\relax
\mciteBstWouldAddEndPuncttrue
\mciteSetBstMidEndSepPunct{\mcitedefaultmidpunct}
{\mcitedefaultendpunct}{\mcitedefaultseppunct}\relax
\EndOfBibitem
\bibitem[von Horsten \latin{et~al.}(2011)von Horsten, Banks, and
  Clary]{claryhalk}
von Horsten,~H.~F.; Banks,~S.~T.; Clary,~D.~C. An Efficient Route to Thermal
  Rate Constants in Reduced Dimensional Quantum Scattering Simulations:
  Applications to the Abstraction of Hydrogen from Alkanes. \emph{J. Chem.
  Phys.} \textbf{2011}, \emph{135}, 094311\relax
\mciteBstWouldAddEndPuncttrue
\mciteSetBstMidEndSepPunct{\mcitedefaultmidpunct}
{\mcitedefaultendpunct}{\mcitedefaultseppunct}\relax
\EndOfBibitem
\bibitem[Marcus and Coltrin(1977)Marcus, and Coltrin]{mc1977}
Marcus,~R.~A.; Coltrin,~M.~E. A New Tunneling Path for Reactions such as
  \uppercase{H+H$_2$} $\rightarrow$ \uppercase{H$_2$+H}. \emph{J. Chem. Phys.}
  \textbf{1977}, \emph{67}, 2609--2613\relax
\mciteBstWouldAddEndPuncttrue
\mciteSetBstMidEndSepPunct{\mcitedefaultmidpunct}
{\mcitedefaultendpunct}{\mcitedefaultseppunct}\relax
\EndOfBibitem
\bibitem[Craig and Manolopoulos(2005)Craig, and Manolopoulos]{rpmd05}
Craig,~I.~R.; Manolopoulos,~D.~E. Chemical Reaction Rates from Ring Polymer
  Molecular Dynamics. \emph{J. Chem. Phys.} \textbf{2005}, \emph{122},
  084106\relax
\mciteBstWouldAddEndPuncttrue
\mciteSetBstMidEndSepPunct{\mcitedefaultmidpunct}
{\mcitedefaultendpunct}{\mcitedefaultseppunct}\relax
\EndOfBibitem
\bibitem[Suleimanov \latin{et~al.}(2013)Suleimanov, de~Tudela, Jambrina,
  Castillo, Sajez-Rabanos, Manolopoulos, and Aoiz]{rpmdh3}
Suleimanov,~Y.~V.; de~Tudela,~R.~P.; Jambrina,~P.~G.; Castillo,~J.~F.;
  Sajez-Rabanos,~V.; Manolopoulos,~D.~E.; Aoiz,~F.~J. A Ring Polymer Molecular
  Dynamics Study of the Isotopologues of the \uppercase{H+H$_2$} Reaction.
  \emph{Phys. Chem. Chem. Phys.} \textbf{2013}, \emph{15}, 3655--3665\relax
\mciteBstWouldAddEndPuncttrue
\mciteSetBstMidEndSepPunct{\mcitedefaultmidpunct}
{\mcitedefaultendpunct}{\mcitedefaultseppunct}\relax
\EndOfBibitem
\bibitem[Richardson(2018)]{rich18}
Richardson,~J.~O. Perspective: Ring-Polymer Instanton Theory. \emph{J. Chem.
  Phys.} \textbf{2018}, \emph{148}, 200901\relax
\mciteBstWouldAddEndPuncttrue
\mciteSetBstMidEndSepPunct{\mcitedefaultmidpunct}
{\mcitedefaultendpunct}{\mcitedefaultseppunct}\relax
\EndOfBibitem
\bibitem[Miller \latin{et~al.}(1983)Miller, Schwartz, and Tromp]{millerexact83}
Miller,~W.~H.; Schwartz,~S.~D.; Tromp,~J.~W. Quantum Mechanical Rate Constants
  for Bimolecular Reactions. \emph{J. Chem. Phys.} \textbf{1983}, \emph{79},
  4889--4898\relax
\mciteBstWouldAddEndPuncttrue
\mciteSetBstMidEndSepPunct{\mcitedefaultmidpunct}
{\mcitedefaultendpunct}{\mcitedefaultseppunct}\relax
\EndOfBibitem
\bibitem[Wu \latin{et~al.}(2004)Wu, Werner, and Manthe]{manthhch4}
Wu,~T.; Werner,~H.-J.; Manthe,~U. First-Principles Theory for the \uppercase{H
  + CH$_4$} $\rightarrow$ \uppercase{H$_2$ + CH$_3$} Reaction. \emph{Science}
  \textbf{2004}, \emph{306}, 2227--2229\relax
\mciteBstWouldAddEndPuncttrue
\mciteSetBstMidEndSepPunct{\mcitedefaultmidpunct}
{\mcitedefaultendpunct}{\mcitedefaultseppunct}\relax
\EndOfBibitem
\bibitem[Welsch and Manthe(2012)Welsch, and Manthe]{manth12}
Welsch,~R.; Manthe,~U. Reaction Dynamics with the Multi-Layer
  Multi-Configurational Time-Dependent Hartree Approach: \uppercase{H + CH$_4$}
  $\rightarrow$ \uppercase{H$_2$ + CH$_3$} Rate Constants for Different
  Potentials. \emph{J. Chem. Phys.} \textbf{2012}, \emph{137}, 244106\relax
\mciteBstWouldAddEndPuncttrue
\mciteSetBstMidEndSepPunct{\mcitedefaultmidpunct}
{\mcitedefaultendpunct}{\mcitedefaultseppunct}\relax
\EndOfBibitem
\bibitem[Clyne and Walker(1973)Clyne, and Walker]{Clyne1973}
Clyne,~M. A.~A.; Walker,~R.~F. Absolute Rate Constants for Elementary Reactions
  in the Chlorination of \ce{CH4}, \ce{CD4}, \ce{CH3Cl}, \ce{CH2Cl2},
  \ce{CHCl3}, \ce{CDCl3} and \ce{CBrCl3}. \emph{J. Chem. Soc., Faraday Trans.
  1.} \textbf{1973}, \emph{69}, 1547--1567\relax
\mciteBstWouldAddEndPuncttrue
\mciteSetBstMidEndSepPunct{\mcitedefaultmidpunct}
{\mcitedefaultendpunct}{\mcitedefaultseppunct}\relax
\EndOfBibitem
\bibitem[Manning and Kurylo(1977)Manning, and Kurylo]{Manning1977}
Manning,~R.~G.; Kurylo,~M.~J. Flash Photolysis Resonance Fluorescence
  Investigation of the Temperature Dependencies of the Reactions of
  Chlorine(2P) Atoms with Methane, Chloromethane, Fluoromethane, Excited
  Fluoromethane, and Ethane. \emph{J. Phys. Chem.} \textbf{1977}, \emph{81},
  291--296\relax
\mciteBstWouldAddEndPuncttrue
\mciteSetBstMidEndSepPunct{\mcitedefaultmidpunct}
{\mcitedefaultendpunct}{\mcitedefaultseppunct}\relax
\EndOfBibitem
\bibitem[Michael and Lee(1977)Michael, and Lee]{Michael1977}
Michael,~J.~V.; Lee,~J.~H. Selected Rate Constants for H, O, N, and Cl Atoms
  with Substrates at Room Temperatures. \emph{Chem. Phys. Lett.} \textbf{1977},
  \emph{51}, 303--306\relax
\mciteBstWouldAddEndPuncttrue
\mciteSetBstMidEndSepPunct{\mcitedefaultmidpunct}
{\mcitedefaultendpunct}{\mcitedefaultseppunct}\relax
\EndOfBibitem
\bibitem[Whytock \latin{et~al.}(1977)Whytock, Lee, Michael, Payne, and
  Stief]{Whytock1977}
Whytock,~D.~A.; Lee,~J.~H.; Michael,~J.~V.; Payne,~W.~A.; Stief,~L.~J. Absolute
  Rate of the Reaction of Cl(2P) with Methane from 200–500 K. \emph{J. Chem.
  Phys.} \textbf{1977}, \emph{66}, 2690--2695\relax
\mciteBstWouldAddEndPuncttrue
\mciteSetBstMidEndSepPunct{\mcitedefaultmidpunct}
{\mcitedefaultendpunct}{\mcitedefaultseppunct}\relax
\EndOfBibitem
\bibitem[Keyser(1978)]{Keyser1978}
Keyser,~L.~F. Absolute Rate and Temperature Dependence of the Reaction between
  Chlorine (2P) Atoms and Methane. \emph{J. Chem. Phys.} \textbf{1978},
  \emph{69}, 214--218\relax
\mciteBstWouldAddEndPuncttrue
\mciteSetBstMidEndSepPunct{\mcitedefaultmidpunct}
{\mcitedefaultendpunct}{\mcitedefaultseppunct}\relax
\EndOfBibitem
\bibitem[Zahniser \latin{et~al.}(1978)Zahniser, Berquist, and
  Kaufman]{Zahniser1978}
Zahniser,~M.~S.; Berquist,~B.~M.; Kaufman,~F. Kinetics of the reaction \ce{Cl +
  CH4} $\rightarrow$ \ce{CH3 + HCl} from 200° to 500° K. \emph{Int. J. Chem.
  Kinet.} \textbf{1978}, \emph{10}, 15--29\relax
\mciteBstWouldAddEndPuncttrue
\mciteSetBstMidEndSepPunct{\mcitedefaultmidpunct}
{\mcitedefaultendpunct}{\mcitedefaultseppunct}\relax
\EndOfBibitem
\bibitem[Ravishankara and Wine(1980)Ravishankara, and Wine]{Ravishankara1980}
Ravishankara,~A.~R.; Wine,~P.~H. A Laser Flash Photolysis‐Resonance
  Fluorescence Kinetics Study of the Reaction \ce{Cl(2P) + CH4} $\rightarrow$
  \ce{CH3 + HCl}. \emph{J. Chem. Phys.} \textbf{1980}, \emph{72}, 25--30\relax
\mciteBstWouldAddEndPuncttrue
\mciteSetBstMidEndSepPunct{\mcitedefaultmidpunct}
{\mcitedefaultendpunct}{\mcitedefaultseppunct}\relax
\EndOfBibitem
\bibitem[Heneghan \latin{et~al.}(1981)Heneghan, Knoot, and
  Benson]{Heneghan1981}
Heneghan,~S.~P.; Knoot,~P.~A.; Benson,~S.~W. The Temperature Coefficient of the
  Rates in the System \ce{Cl + CH4} = \ce{CH3 + HCl}, Thermochemistry of the
  Methyl Radical. \emph{Int. J. Chem. Kinet.} \textbf{1981}, \emph{13},
  677--691\relax
\mciteBstWouldAddEndPuncttrue
\mciteSetBstMidEndSepPunct{\mcitedefaultmidpunct}
{\mcitedefaultendpunct}{\mcitedefaultseppunct}\relax
\EndOfBibitem
\bibitem[Seeley \latin{et~al.}(1996)Seeley, Jayne, and Molina]{Seeley1996}
Seeley,~J.~V.; Jayne,~J.~T.; Molina,~M.~J. Kinetic Studies of Chlorine Atom
  Reactions Using the Turbulent Flow Tube Technique. \emph{J. Phys. Chem.}
  \textbf{1996}, \emph{100}, 4019--4025\relax
\mciteBstWouldAddEndPuncttrue
\mciteSetBstMidEndSepPunct{\mcitedefaultmidpunct}
{\mcitedefaultendpunct}{\mcitedefaultseppunct}\relax
\EndOfBibitem
\bibitem[Pilgrim \latin{et~al.}(1997)Pilgrim, McIlroy, and
  Taatjes]{Pilgrim1997}
Pilgrim,~J.~S.; McIlroy,~A.; Taatjes,~C.~A. Kinetics of Cl Atom Reactions with
  Methane, Ethane, and Propane from 292 to 800 K. \emph{J. Chem. Phys. A.}
  \textbf{1997}, \emph{101}, 1873--1880\relax
\mciteBstWouldAddEndPuncttrue
\mciteSetBstMidEndSepPunct{\mcitedefaultmidpunct}
{\mcitedefaultendpunct}{\mcitedefaultseppunct}\relax
\EndOfBibitem
\bibitem[Bryukov \latin{et~al.}(2002)Bryukov, Slagle, and Knyazev]{Bryukov2002}
Bryukov,~M.~G.; Slagle,~I.~R.; Knyazev,~V.~D. Kinetics of Reactions of Cl Atoms
  with Methane and Chlorinated Methanes. \emph{J. Chem. Phys. A.}
  \textbf{2002}, \emph{106}, 10532--10542\relax
\mciteBstWouldAddEndPuncttrue
\mciteSetBstMidEndSepPunct{\mcitedefaultmidpunct}
{\mcitedefaultendpunct}{\mcitedefaultseppunct}\relax
\EndOfBibitem
\bibitem[Czakó and Bowman~Joel(2011)Czakó, and Bowman~Joel]{CzakoBowman2011}
Czakó,~G.; Bowman~Joel,~M. Dynamics of the Reaction of Methane with Chlorine
  Atom on an Accurate Potential Energy Surface. \emph{Science} \textbf{2011},
  \emph{334}, 343--346\relax
\mciteBstWouldAddEndPuncttrue
\mciteSetBstMidEndSepPunct{\mcitedefaultmidpunct}
{\mcitedefaultendpunct}{\mcitedefaultseppunct}\relax
\EndOfBibitem
\bibitem[Liu and Li(2020)Liu, and Li]{LL2020}
Liu,~Y.; Li,~J. An Accurate Potential Energy Surface and Ring Polymer Molecular
  Dynamics Study of the Cl + \ce{CH4} → HCl + \ce{CH3} Reaction. \emph{Phys.
  Chem. Chem. Phys.} \textbf{2020}, \emph{22}, 344--353\relax
\mciteBstWouldAddEndPuncttrue
\mciteSetBstMidEndSepPunct{\mcitedefaultmidpunct}
{\mcitedefaultendpunct}{\mcitedefaultseppunct}\relax
\EndOfBibitem
\bibitem[Barker \latin{et~al.}(2012)Barker, Nguyen, and
  Stanton]{BarkerNguyenStanton2012}
Barker,~J.~R.; Nguyen,~T.~L.; Stanton,~J.~F. Kinetic Isotope Effects for Cl +
  \ce{CH4} = HCl + \ce{CH3} Calculated Using ab Initio Semiclassical Transition
  State Theory. \emph{J. Chem. Phys. A.} \textbf{2012}, \emph{116},
  6408--6419\relax
\mciteBstWouldAddEndPuncttrue
\mciteSetBstMidEndSepPunct{\mcitedefaultmidpunct}
{\mcitedefaultendpunct}{\mcitedefaultseppunct}\relax
\EndOfBibitem
\bibitem[Georgievskii and Klippenstein(2021)Georgievskii, and
  Klippenstein]{GK2021}
Georgievskii,~Y.; Klippenstein,~S.~J. Entanglement Effect and Angular Momentum
  Conservation in a Nonseparable Tunneling Treatment. \emph{J. Chem. Theory
  Comput.} \textbf{2021}, \emph{17}, 3863--3885\relax
\mciteBstWouldAddEndPuncttrue
\mciteSetBstMidEndSepPunct{\mcitedefaultmidpunct}
{\mcitedefaultendpunct}{\mcitedefaultseppunct}\relax
\EndOfBibitem
\bibitem[Hoppe and Manthe(2022)Hoppe, and Manthe]{HM2022}
Hoppe,~H.; Manthe,~U. First-Principles Theory for the Reaction of Chlorine with
  Methane. \emph{J. Phys. Chem. Lett.} \textbf{2022}, \emph{13},
  2563--2566\relax
\mciteBstWouldAddEndPuncttrue
\mciteSetBstMidEndSepPunct{\mcitedefaultmidpunct}
{\mcitedefaultendpunct}{\mcitedefaultseppunct}\relax
\EndOfBibitem
\bibitem[Houston \latin{et~al.}(2019)Houston, Nandi, and Bowman]{houston19}
Houston,~P.~L.; Nandi,~A.; Bowman,~J.~M. A Machine Learning Approach for
  Prediction of Rate Constants. \emph{J. Phys. Chem. Letts} \textbf{2019},
  \emph{10}, 5250--5258\relax
\mciteBstWouldAddEndPuncttrue
\mciteSetBstMidEndSepPunct{\mcitedefaultmidpunct}
{\mcitedefaultendpunct}{\mcitedefaultseppunct}\relax
\EndOfBibitem
\bibitem[Nandi \latin{et~al.}(2020)Nandi, Bowman, and Houston]{NandiOHCl2020}
Nandi,~A.; Bowman,~J.~M.; Houston,~P. A Machine Learning Approach for Rate
  Constants. II. Clustering, Training, and Predictions for the O(3P) + HCl →
  OH + Cl Reaction. \emph{J. Phys. Chem. A.} \textbf{2020}, \emph{124},
  5746--5755\relax
\mciteBstWouldAddEndPuncttrue
\mciteSetBstMidEndSepPunct{\mcitedefaultmidpunct}
{\mcitedefaultendpunct}{\mcitedefaultseppunct}\relax
\EndOfBibitem
\bibitem[Komp and Valleau(2020)Komp, and Valleau]{valleau2020}
Komp,~E.; Valleau,~S. Machine Learning Quantum Reaction Rate Constants.
  \emph{J. Phys. Chem. A} \textbf{2020}, \emph{124}, 8607--8613\relax
\mciteBstWouldAddEndPuncttrue
\mciteSetBstMidEndSepPunct{\mcitedefaultmidpunct}
{\mcitedefaultendpunct}{\mcitedefaultseppunct}\relax
\EndOfBibitem
\bibitem[Komp \latin{et~al.}(2021)Komp, Janulaitis, and Valleau]{Valleau2021}
Komp,~E.; Janulaitis,~N.; Valleau,~S. Progress towards Machine Learning
  Reaction Rate Constants. \emph{Phys. Chem. Chem. Phys.} \textbf{2021}, \relax
\mciteBstWouldAddEndPunctfalse
\mciteSetBstMidEndSepPunct{\mcitedefaultmidpunct}
{}{\mcitedefaultseppunct}\relax
\EndOfBibitem
\bibitem[Lewis-Atwell \latin{et~al.}(2021)Lewis-Atwell, Townsend, and
  Grayson]{Grayson2021}
Lewis-Atwell,~T.; Townsend,~P.~A.; Grayson,~M.~N. Machine Learning Activation
  Energies of Chemical Reactions. \emph{WIREs. Comput. Mole. Sci.}
  \textbf{2021}, \emph{n/a}, e1593\relax
\mciteBstWouldAddEndPuncttrue
\mciteSetBstMidEndSepPunct{\mcitedefaultmidpunct}
{\mcitedefaultendpunct}{\mcitedefaultseppunct}\relax
\EndOfBibitem
\bibitem[Green(2019)]{Green2019}
Green,~W.~H. In \emph{Computer Aided Chemical Engineering}; Faravelli,~T.,
  Manenti,~F., Ranzi,~E., Eds.; Elsevier, 2019; Vol.~45; pp 259--294\relax
\mciteBstWouldAddEndPuncttrue
\mciteSetBstMidEndSepPunct{\mcitedefaultmidpunct}
{\mcitedefaultendpunct}{\mcitedefaultseppunct}\relax
\EndOfBibitem
\bibitem[Grambow \latin{et~al.}(2020)Grambow, Pattanaik, and Green]{Green2020}
Grambow,~C.~A.; Pattanaik,~L.; Green,~W.~H. Deep Learning of Activation
  Energies. \emph{J. Phys. Chem. Lett.} \textbf{2020}, \emph{11},
  2992--2997\relax
\mciteBstWouldAddEndPuncttrue
\mciteSetBstMidEndSepPunct{\mcitedefaultmidpunct}
{\mcitedefaultendpunct}{\mcitedefaultseppunct}\relax
\EndOfBibitem
\bibitem[Brown(1981)]{Brown1981}
Brown,~R.~L. A Method of Calculating Tunneling Corrections for Eckart Potential
  Barriers. \emph{J. Res. Nat. Bur. Stand.} \textbf{1981}, \emph{86},
  357--359\relax
\mciteBstWouldAddEndPuncttrue
\mciteSetBstMidEndSepPunct{\mcitedefaultmidpunct}
{\mcitedefaultendpunct}{\mcitedefaultseppunct}\relax
\EndOfBibitem
\bibitem[Rasmussen and Williams(2006)Rasmussen, and Williams]{Rasmussen2004}
Rasmussen,~C.~E.; Williams,~C. K.~I. \emph{Gaussian processes for machine
  learning}; the MIT Press, 2006\relax
\mciteBstWouldAddEndPuncttrue
\mciteSetBstMidEndSepPunct{\mcitedefaultmidpunct}
{\mcitedefaultendpunct}{\mcitedefaultseppunct}\relax
\EndOfBibitem
\bibitem[Vieira and Krems(2017)Vieira, and Krems]{krems17}
Vieira,~D.; Krems,~R.~V. Rate Constants for Fine-structure Excitations in
  \uppercase{O}{\textendash}\uppercase{H} Collisions with Error Bars Obtained
  by Machine Learning. \emph{Astrophys. J.} \textbf{2017}, \emph{835},
  255\relax
\mciteBstWouldAddEndPuncttrue
\mciteSetBstMidEndSepPunct{\mcitedefaultmidpunct}
{\mcitedefaultendpunct}{\mcitedefaultseppunct}\relax
\EndOfBibitem
\bibitem[Bart\'ok and Cs\'anyi(2015)Bart\'ok, and Cs\'anyi]{GP-2015-1}
Bart\'ok,~A.~P.; Cs\'anyi,~G. Gaussian Approximation Potentials: A Brief
  Tutorial Introduction. \emph{Int. J. Quantum Chem.} \textbf{2015},
  \emph{115}, 1051--1057\relax
\mciteBstWouldAddEndPuncttrue
\mciteSetBstMidEndSepPunct{\mcitedefaultmidpunct}
{\mcitedefaultendpunct}{\mcitedefaultseppunct}\relax
\EndOfBibitem
\bibitem[Cui and Krems(2016)Cui, and Krems]{Cui2016}
Cui,~J.; Krems,~R.~V. Efficient Non-Parametric Fitting of Potential Energy
  Surfaces for Polyatomic Molecules with Gaussian Processes. \emph{J. Phys. B:
  At. Mol. Opt. Phys.} \textbf{2016}, \emph{49}\relax
\mciteBstWouldAddEndPuncttrue
\mciteSetBstMidEndSepPunct{\mcitedefaultmidpunct}
{\mcitedefaultendpunct}{\mcitedefaultseppunct}\relax
\EndOfBibitem
\bibitem[Pedregosa \latin{et~al.}(2011)Pedregosa, Varoquaux, Gramfort, Michel,
  Thirion, Grisel, Blondel, Prettenhofer, Weiss, Dubourg, Vanderplas, Passos,
  Cournapeau, Brucher, Perrot, and Duchesnay]{scikit-learn}
Pedregosa,~F.; Varoquaux,~G.; Gramfort,~A.; Michel,~V.; Thirion,~B.;
  Grisel,~O.; Blondel,~M.; Prettenhofer,~P.; Weiss,~R.; Dubourg,~V.
  \latin{et~al.}  Scikit-learn: Machine Learning in {P}ython. \emph{J. Mach.
  Learn. Res.} \textbf{2011}, \emph{12}, 2825--2830\relax
\mciteBstWouldAddEndPuncttrue
\mciteSetBstMidEndSepPunct{\mcitedefaultmidpunct}
{\mcitedefaultendpunct}{\mcitedefaultseppunct}\relax
\EndOfBibitem
\bibitem[Meng \latin{et~al.}(2015)Meng, Chen, and Zhang]{rpmdch52015}
Meng,~Q.; Chen,~J.; Zhang,~D.~H. Communication: Rate Coefficients of the
  \uppercase{H + CH$_4$} $\rightarrow$ \uppercase{H$_2$ + CH$_3$} Reaction from
  Ring Polymer Molecular Dynamics on a Highly Accurate Potential Energy
  Surface. \emph{J. Chem. Phys.} \textbf{2015}, \emph{143}, 101102\relax
\mciteBstWouldAddEndPuncttrue
\mciteSetBstMidEndSepPunct{\mcitedefaultmidpunct}
{\mcitedefaultendpunct}{\mcitedefaultseppunct}\relax
\EndOfBibitem
\bibitem[Li \latin{et~al.}(2014)Li, Suleimanov, Green, and Guo]{rpmdclch4}
Li,~Y.; Suleimanov,~Y.~V.; Green,~W.~H.; Guo,~H. Quantum Rate Coefficients and
  Kinetic Isotope Effect for the Reaction \uppercase{C}l + \uppercase{CH$_4$}
  $\rightarrow$ \uppercase{HC}l + \uppercase{CH$_3$} from Ring Polymer
  Molecular Dynamics. \emph{J. Phys. Chem. A.} \textbf{2014}, \emph{118},
  1989--1996\relax
\mciteBstWouldAddEndPuncttrue
\mciteSetBstMidEndSepPunct{\mcitedefaultmidpunct}
{\mcitedefaultendpunct}{\mcitedefaultseppunct}\relax
\EndOfBibitem
\bibitem[Menendez \latin{et~al.}(2019)Menendez, Jambrina, Zanchet, Verdasco,
  Suleimanov, and Aoiz]{rpmd2019}
Menendez,~M.; Jambrina,~P.~G.; Zanchet,~A.; Verdasco,~E.; Suleimanov,~Y.~V.;
  Aoiz,~F.~J. New Stress Test for Ring Polymer Molecular Dynamics: Rate
  Coefficients of the \uppercase{O($^3$P)} + \uppercase{HC}l Reaction and
  Comparison with Quantum Mechanical and Quasiclassical Trajectory Results.
  \emph{J. Phys. Chem. A.} \textbf{2019}, \emph{123}, 7920--7931\relax
\mciteBstWouldAddEndPuncttrue
\mciteSetBstMidEndSepPunct{\mcitedefaultmidpunct}
{\mcitedefaultendpunct}{\mcitedefaultseppunct}\relax
\EndOfBibitem
\bibitem[Xie \latin{et~al.}(2002)Xie, Wang, Bowman, and Manolopoulos]{tiao2002}
Xie,~T.; Wang,~D.; Bowman,~J.~M.; Manolopoulos,~D.~E. Resonances in the
  O($^3$P)+HCl Reaction Due to Van der Waals Minima. \emph{J. Chem. Phys.}
  \textbf{2002}, \emph{116}, 7461--7467\relax
\mciteBstWouldAddEndPuncttrue
\mciteSetBstMidEndSepPunct{\mcitedefaultmidpunct}
{\mcitedefaultendpunct}{\mcitedefaultseppunct}\relax
\EndOfBibitem
\bibitem[Xie \latin{et~al.}(2003)Xie, Bowman, Peterson, and
  Ramachandran]{xieohcl}
Xie,~T.; Bowman,~J.~M.; Peterson,~K.~A.; Ramachandran,~B. Quantum Calculations
  of the Rate Constant for the \uppercase{O($^3$P)+HC}l Reaction on New Ab
  initio \uppercase{A$'$} and \uppercase{A$''$} Surfaces. \emph{J. Chem. Phys.}
  \textbf{2003}, \emph{119}, 9601--9608\relax
\mciteBstWouldAddEndPuncttrue
\mciteSetBstMidEndSepPunct{\mcitedefaultmidpunct}
{\mcitedefaultendpunct}{\mcitedefaultseppunct}\relax
\EndOfBibitem
\bibitem[Bowman(2005)]{bowman05}
Bowman,~J.~M. Enhancement of Tunneling Due to Resonances in Pre-barrier Wells
  in Chemical Reactions. \emph{Chem. Phys.} \textbf{2005}, \emph{308}, 255 --
  257\relax
\mciteBstWouldAddEndPuncttrue
\mciteSetBstMidEndSepPunct{\mcitedefaultmidpunct}
{\mcitedefaultendpunct}{\mcitedefaultseppunct}\relax
\EndOfBibitem
\end{mcitethebibliography}

\newpage
\includegraphics[width=\textwidth,page=1]{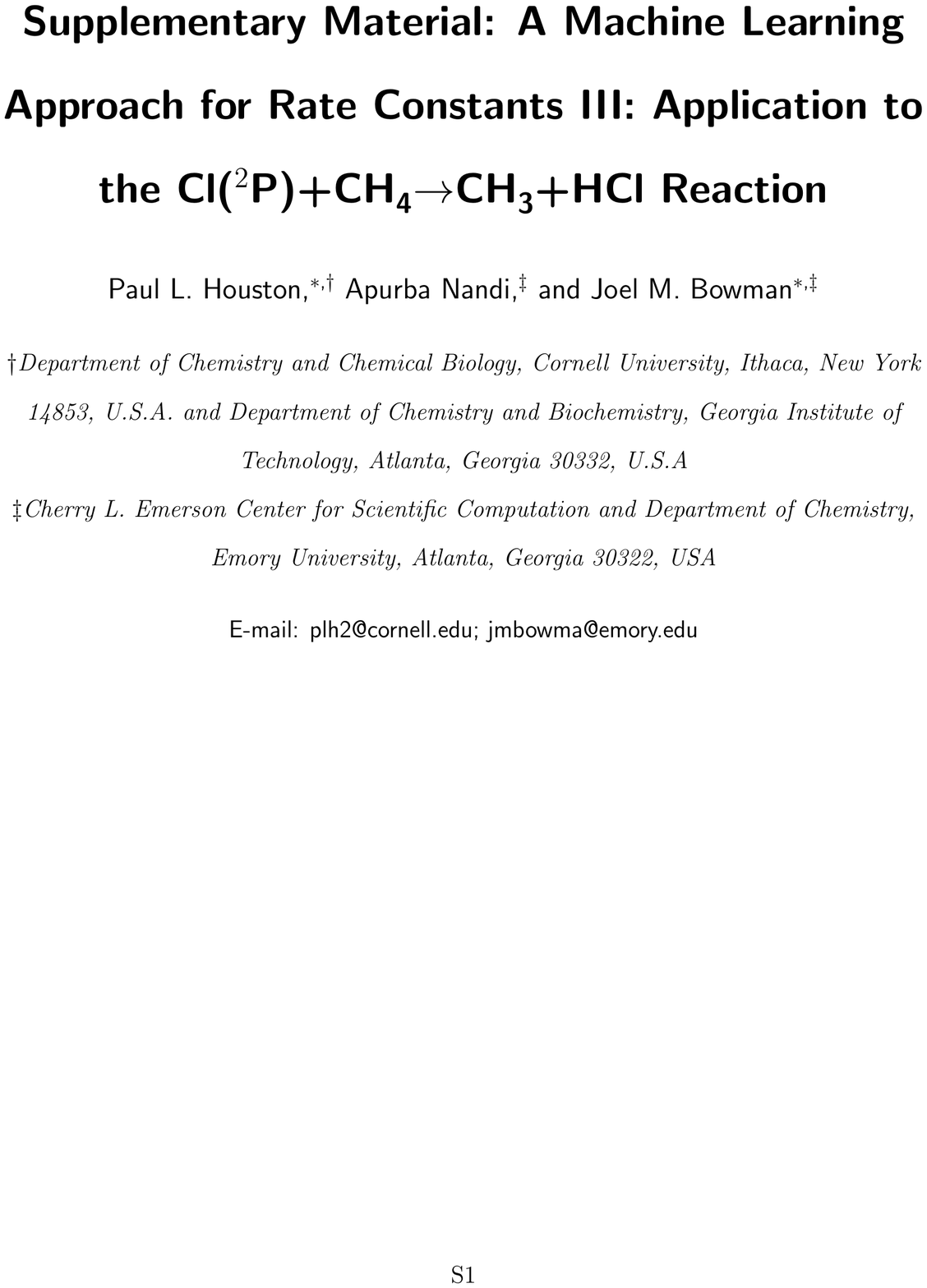}
\includegraphics[width=\textwidth,page=2]{SI.pdf}
\includegraphics[width=\textwidth,page=3]{SI.pdf}
\includegraphics[width=\textwidth,page=4]{SI.pdf}
\includegraphics[width=\textwidth,page=5]{SI.pdf}
\includegraphics[width=\textwidth,page=6]{SI.pdf}
\includegraphics[width=\textwidth,page=7]{SI.pdf}
\includegraphics[width=\textwidth,page=8]{SI.pdf}
\includegraphics[width=\textwidth,page=9]{SI.pdf}
\end{document}